\documentclass[aps,prl,reprint,twocolumn,nofootinbib]{revtex4}

\usepackage{epsfig,amssymb,amsmath}

\usepackage{color}

\def\erf{\mbox{erf}}

\newcommand{\ket}[1]{|#1\rangle}
\newcommand{\bra}[1]{\langle #1|}
\newcommand{\ketbra}[2]{|#1\rangle\langle #2|}

\def\comment#1{ [{\bf Comment:} {\sf #1}]}
\def\labell#1{\label{#1}}
\def\>{\rangle}\def\<{\langle}
\def\sectiona#1{{\par\em #1:--- }}
\def\togli#1{}

\begin{document}


\title{Digital Quantum Estimation} \author{Lorenzo Maccone$^{1,*}$,
  Majid Hassani$^2$, Chiara Macchiavello$^1$}
\affiliation{\vbox{$^1$Dip.~Fisica and INFN Sez.~Pavia, University~of
    Pavia, via Bassi 6, I-27100 Pavia, Italy} \\ \vbox{$^2$Department
    of Physics, Sharif University of Technology, Tehran 14588,
    Iran}\\$^*$Corresponding Author: maccone@unipv.it}
\begin{abstract}
  { Quantum Metrology calculates the ultimate precision of all
    estimation strategies, measuring what is their root mean-square
    error (RMSE) and their Fisher information. Here, instead, we ask
    {\it how many bits} of the parameter we can recover, namely we
    derive an information-theoretic quantum metrology. In this setting
    we redefine ``Heisenberg bound'' and ``standard quantum limit''
    (the usual benchmarks in quantum estimation theory), and show that
    the former can be attained only by sequential strategies or
    parallel strategies that employ entanglement among probes, whereas
    parallel-separable strategies are limited by the latter.  We
    highlight the differences between this setting and the RMSE-based
    one.}
\end{abstract}


\maketitle  

The theory of quantum metrology
\cite{kok,caves,review,qmetr,matteo,rafalreview,geza,rafalguta,walm,davidov,book,gabriel}
determines the ultimate precision in any estimation.  The estimation
of an unknown parameter generally requires a probe that interacts with
the system to be sampled: the interaction encodes the parameter onto
the probe, which is then measured.\togli{E.g.~to measure the length of
  a table, one prepares a ruler (the probe), places it close to the
  table (interaction), and looks at the markings (probe measurement).}
Clearly, if one uses $N$ independent measurements, the root mean
square error (RMSE) in the estimation scales as $1/\sqrt{N}$ (the
standard quantum limit) as dictated by the central limit theorem.  If
one uses $N$ parallel entangled probes or one probe sequentially $N$
times, the error can be reduced to $1/N$ (the Heisenberg bound)
\cite{qmetr,entangl}. This precision can be attained without the use
of entanglement at the measurement stage \cite{qmetr}.

The RMSE is, however, ill suited for digital sensors, digital data
processing, or even for the digital archival of parameters, where the
number of significant digits (bits) is a more useful figure of merit.
Moreover, the techniques used in the conventional theory (e.g.~the use
of N00N states \cite{dowling}) suffer from ambiguities in the typical
case in which a phase is estimated \cite{bartlett,njp}, so that the
reported RMSE does not typically refer to the true total error in the
estimation \cite{wiseman,morgan,combes}.

In this paper we overcome these problems by replacing RMSE (and Fisher
information) with mutual information, which directly measures the
number of bits of information that the quantum estimation strategy
provides. Namely, we derive an information-theoretic quantum
metrology, obtaining a number of results: (1)~we redefine in a natural
way the concepts of Heisenberg bound (using the Holevo theorem) and of
standard quantum limit; (2)~for parallel estimation strategies the
Heisenberg bound can be attained, but only in the presence of
entanglement, as in the RMSE case; (3)~as expected, for parallel
strategies without entanglement at the preparation, at most the
standard quantum limit is achievable (and entanglement at the
measurement stage is useless); (4)~instead, for sequential strategies
(where one of the probes performs most of the samplings) the
Heisenberg bound is attainable without using entanglement, as in the
RMSE case; (5)~increasing the Hilbert space dimension of the probe is
helpful, in contrast to the RMSE case where a two-dimensional subspace
is sufficient; (6)~the Heisenberg bound is achieved by the quantum
phase estimation algorithm (QPEA) \cite{qpea,qpea2} and by the
Pegg-Barnett phase states \cite{pb}, in contrast to the RMSE case
\cite{wiseman,morgan,nt}.

\sectiona{Heisenberg bound and standard quantum limit}
In quantum metrology we estimate a parameter $\varphi$ by first
preparing one or more probes into an initial state $\rho_0$, then
evolving them by applying $N$ times the interaction $U_\varphi$ that
encodes the parameter onto the probe(s) and transforms the state into
$\rho_\varphi$, and finally measuring $\rho_\varphi$.  The aim is to
find the ultimate precision attainable for the estimation strategy as
a function of $N$. If the probe is finite-dimensional, no estimation
strategy can beat the Heisenberg bound $\propto1/N$ for the RMSE.  

A natural way to extend the Heisenberg bound to an
information-theoretic setting is to use the Holevo theorem
\cite{holevobook}, which gives the maximum number of bits $I$
attainable on a parameter $\varphi$ encoded into a state
$\rho_\varphi$, given the measurement results $\vec m$: \begin{align}
  I(\vec m:\varphi)\leqslant S(\sum_\varphi
  p_\varphi\rho_\varphi)-\sum_\varphi p_\varphi S(\rho_\varphi)
\labell{holevo}\;,
\end{align}
where $S(\rho)=-$Tr$[\rho\log_2\rho]$ is the von Neumann entropy, and
$p_\varphi$ the prior probability of the parameter $\varphi$.  Clearly
the accessible information is largest when $\rho_\varphi$ are all pure
states, and in this case the last sum in \eqref{holevo} is null and
the Holevo bound is attainable. We then define the info-theoretic
Heisenberg bound as $S(\sum_\varphi p_\varphi\rho_\varphi)$. This
quantity scales as $\log_2N$ since we are applying $N$ times the {\em
  same} transformation $U_\varphi$ that encodes the unknown parameter
$\varphi$ (supplementary material). So the Heisenberg bound is
$I\simeq\log_2N$, at least asymptotically for large $N$.  In the RMSE
case the best precision attainable for unentangled parallel strategies
scales as the square root of the Heisenberg bound, so an intuitive
definition of information-theoretic standard quantum limit is
$I\simeq\log_2\sqrt{N}=\tfrac 12\log_2N$. As shown below, this is the
correct definition since unentangled parallel strategies are indeed
bounded by this quantity. These definitions are consistent with the
RMSE based ones: an error $\Delta\varphi\simeq 1/N$ leads to the
expectation that roughly $\log_2N$ binary digits of the results are
reliable, and similarly an error $\Delta\varphi\simeq 1/\sqrt{N}$
leads to the expectation that $\tfrac12\log_2N$ digits are reliable.
Nonetheless, the RMSE and the mutual information capture different
aspects of the estimation's quality, as shown below.

Below we show which kinds of estimation strategies achieve these
bounds.  An example (the QPEA) shows that sequential and
entangled-parallel strategies can achieve the info-theoretic
Heisenberg bound. We then show that the optimal parallel-separable
strategies can only attain the standard quantum limit. We finally
discuss the role of the probe's dimensionality.

\subsection*{Methods}
For the sake of simplicity we will first
restrict to two-dimensional probes (qubits), for which
$U_\varphi=|0\>\<0|+e^{i2\pi \varphi}|1\>\<1|$ (with $|0\>$ and $|1\>$
the eigenstates of the generator of $U_\varphi$), and then separately
analyze what happens in the (finite) $d$-dimension case. We use
finite-dimensional probes and unitaries, so the parameter $\varphi$ is
periodic and we restrict to $\varphi\in[0,1]$. As is customary in
quantum metrology, we request no prior knowledge on the parameter to
be estimated (uniform prior).

\begin{figure}[hbt]
\begin{center}
\epsfxsize=.8\hsize\leavevmode\epsffile{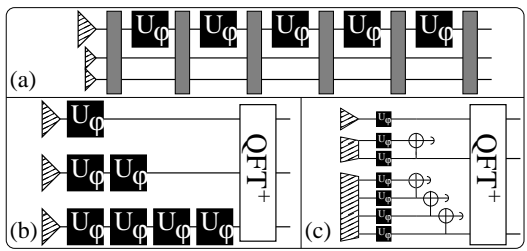}
\end{center}
\vspace{-.5cm}
\caption{Sequential and parallel-entangled strategies.  (a)~Sequential
  strategy, where a single probe (large triangle) samples $N$
  unitaries $U_\varphi$ (black boxes) sequentially.  Ancillary systems
  (small triangles) may interact through arbitrary intermediate
  unitaries (gray squares).  (b)~Quantum phase estimation algorithm
  (QPEA). To see that it is equivalent to a sequential strategy
  \cite{qpea2}, where the last unitary is the inverse quantum Fourier
  transform (QFT$^\dag$), use intermediate unitaries that swap the
  state of the ancillas with the state of the probe.  The output
  (measured in the computational basis) is a ${t}$-bit digital
  estimate of the parameter $\varphi$ with ${t}=\log_2(N+1)$.
  (c)~Parallel QPEA, which uses entangled N00N states (dashed boxes)
  composed of $1,2,4,\cdots, 2^{{t}-1}$ qubits.  The circles represent
  C-NOT gates that remove the entanglement, and the cups represent the
  discarding of qubits in the state $|0\>$.}
\label{f:qpea}\end{figure}

\sectiona{Sequential strategies}
In sequential strategies \cite{qmetr,nature,qpea2} the transformations
$U_\varphi$ act on a single probe sequentially and ancillas may
interact with the probe at any intermediate stage, Fig.~\ref{f:qpea}a.
We consider the QPEA \cite{qpea,qpea2} as an example of sequential
strategy, Fig.~\ref{f:qpea}b: it needs ${t}=\log_2(N+1)$ qubits
initialized in $|+\>\propto|0\>+|1\>$ states, where the zero-th qubit
is subject to $U_\varphi$ once, and the $j$-th qubit is subject to
$U_\varphi$ $2^j$ times. The $t$ qubits then undergo a quantum Fourier
transform (QFT) and are measured in the computational basis, yielding
a $t$-bit number $m$, from which $\varphi$ can be estimated as
$m/2^t$. One can see that the QPEA is a sequential strategy by
considering one of the qubits as the probe and the others as ancillas,
and inserting appropriate swap-unitaries to swap the ancilla states
and the probe state (the zero-th swap after a single $U_\varphi$
action, the $j$-th after $2^j$ actions) \cite{qpea2}.  

\begin{figure}[hbt]
\begin{center}
\epsfxsize=.95\hsize\leavevmode\epsffile{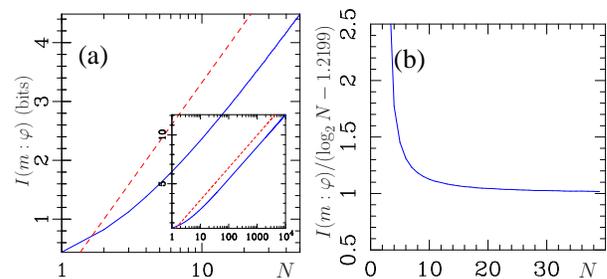}
\end{center}
\vspace{-.5cm}
\caption{Heisenberg bound of the QPEA. (a)~Plot of the mutual
  information $I(m:\varphi)$ as a function of $N$ (blue) and of the
  function $\log_2N$ (dashed red).  Note that $I$ quickly acquires the
  same linear dependence in a log scale as the Heisenberg bound. The
  inset shows the same behavior for large $N$. (b)~Ratio between the
  mutual information and $\log_2N-1.2199$, showing the rapid onset of
  the asymptotic behavior to this quantity.}
\label{f:mutinf}\end{figure}

To evaluate how many of the bits of $m$ are reliable, one needs to
calculate the mutual information $I(m:\varphi)$, using the QPEA
conditional probability
\begin{align}
p(m|\varphi)=\frac{\sin^2(\pi (N+1)
    \varphi)}{(N+1)^2\sin^2(\pi(\varphi-m/(N+1)))}
\labell{probqpea}\;.
\end{align}
The mutual information obtained from it has an asymptotic scaling in
$N$ given by (supplementary material)
\begin{align}
I(m:\varphi)\to \log_2N-2+2\tfrac{\gamma+\ln(2)-1}{\ln(2)}\simeq 
\log_2N-1.2199,\nonumber\\
\labell{mmm}
\end{align}
where $\gamma$ is the Euler-Mascheroni constant. Namely, it (quickly)
asymptotically achieves the info-theoretic Heisenberg bound, apart
from a small additive constant, see Fig.~\ref{f:mutinf}.

The QPEA is known to also achieve the best estimation in terms of a
window function cost \cite{qpea2}, but it cannot achieve the
RMSE-based Heisenberg bound unless one repeats it a few times
\cite{wiseman,morgan,combes}.

\sectiona{Parallel entangled strategies}
The proof that parallel entangled strategies can achieve the
mutual-info Heisenberg bound is simple, since one can easily transform
the sequential strategy detailed above into a parallel one by
entangling the probes: see Fig.~\ref{f:qpea}c. This means that one
uses $N$ probes grouped in N00N states of increasing number of bits:
$|0\>+|1\>$, $|00\>+|11\>,\cdots,$ $|0\>^{2^j}+|1\>^{2^j},\cdots$.
When these $\log_2(N+1)$ groups interact in parallel with the
$N$ transformations $U_\varphi$, the $j$th group acquires a phase of
$2\pi 2^j\varphi$, the same as the corresponding probe in the QPEA strategy
of Fig.~\ref{f:qpea}b. A simple network of controlled-not gates can
transfer this phase to one of the probes in each group and the other
probes in the group are discarded. So the input to the final quantum
Fourier transform is identical to the one of the conventional QPEA.
Thus both the output probability and the mutual information are the
same as the ones calculated above: also the parallel entangled strategy
can achieve the Heisenberg bound (apart from a small additive
constant).

Note that the use of controlled-not gates after the action of the
transformations $U_\varphi$ imply that this procedure requires an
entangled detection strategy (in contrast, the QFT does not require
entanglement among probes \cite{qft}). It is still an open question
whether a parallel entangled strategy can achieve the info-theoretic
Heisenberg bound with a separable detection, as is the case for the
RMSE bound.  The Heisenberg bound is {\em not} achieved (supplementary
material) if one uses the same detection strategy as in the RMSE case
(namely projecting each probe onto the $|\pm\>\propto|0\>\pm|1\>$
states) or if one employs the single-qubit optimal strategy according
to Davies theorem (see below).

\sectiona{Parallel separable strategies}
To prove that without entanglement the parallel strategies cannot
achieve the Heisenberg bound, one needs to analyze the optimal
strategy and show that it can only achieve the standard quantum limit.
(Whereas to prove that the sequential and entangled strategies can
achieve the Heisenberg bound, we merely had to exhibit an example, the
QPEA above.) \togli{As shown below, the optimal strategy requires entangled
measurements, but we will also present a strategy that uses separable
measurements and can still achieve the standard quantum limit.}

In the separable case, the optimal input state for each qubit
probe is an equatorial state, such as $(|0\>+|1\>)/\sqrt{2}$, which is
evolved by $U_\varphi$ into
$|\varphi\>=(|0\>+e^{i2\pi \varphi}|1\>)/\sqrt{2}$. Indeed equatorial
states maximize the distinguishability between input and output. The
$N$ parallel probes after the $U_\varphi$ evolutions emerge in a joint
state
\begin{eqnarray}
|\varphi\>^{\otimes N}=\sum_{j=0}^N\sqrt{\frac
  1{2^N}\left(\begin{matrix}N\cr j\end{matrix}\right)}e^{i2\pi j\varphi
}|S_j\>, 
\labell{inpst}\;
\end{eqnarray}
where $|S_j\>$ is the normalized symmetric state obtained by summing
over all possible permutations with $j$ ones, e.g.~for $N=4$,
$|S_1\>\propto|0001\>+|0010\>+|0100\>+|1000\>$,
$|S_2\>\propto|0011\>+|0101\>+|0110\>+|1001\>+|1010\>+|1100\>$.

To obtain the POVM that maximizes the mutual information on this
state, we use Davies' theorem \cite{davies}: If the input is covariant
with respect to a group that admits an irreducible unitary
representation $U_\varphi$, then there exists a unit vector $|r\>$
such that the mutual information is maximized by the positive
operator-valued measure (POVM)
\begin{eqnarray}
  \Pi_\phi=\tfrac d{|G|}U_\phi|r\>\<r|U^\dag_\phi
\labell{povm}\;,
\end{eqnarray}
where $d$ is the dimension of the system Hilbert space and $|G|$ is
the number of elements in the group \cite{davies}. Davies' theorem can
be extended to continuous parameters $\varphi$ by requiring the
compactness of the group \cite{giuliophd} and to unitary
representations that are irreducible only on equatorial states
\cite{sasaki}.

\begin{figure}[hbt]
\begin{center}
\epsfxsize=.95\hsize\leavevmode\epsffile{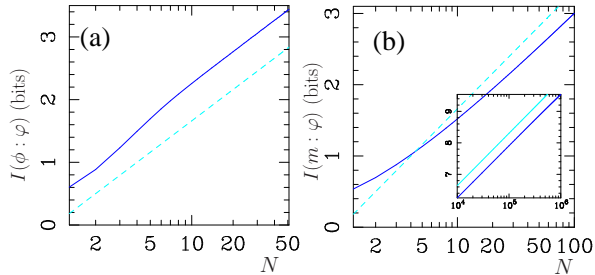}
\end{center}
\vspace{-.5cm}
\caption{Standard quantum limit for unentangled probes. (a)~Plot of
  the mutual information $I(\phi:\varphi)$ relative to the optimal
  POVM \eqref{povmo}, which uses entangled measurements, as a function
  of $N$ (blue), and of the standard quantum limit $\log_2(N)/2$ (cyan
  dashed).  The fluctuations are due to the Monte-Carlo integration
  used here. (b)~Plot of the mutual information $I(\vec m:\varphi)$ of
  \eqref{mt}, relative to the separable POVM that projects onto
  $|\pm\>$ each probe, as a function of $N$ (blue), and of the
  standard quantum limit $\log_2(N)/2$ (cyan dashed). The inset shows
  the large-$N$ scaling. Both cases asymptotically scale at the
  standard quantum limit (apart from a small additive constant).  }
\label{f:numeric}\end{figure}

Since the state $|\varphi\>^{\otimes N}$ spans only the
$N+1$-dimensional symmetric subspace of the $N$-qubits space, we can
limit ourselves to it. So the optimal POVM is given by \eqref{povm}
with $d=N+1$, $|G|=1$ and $|r\>$ a state in the symmetric subspace:
$|r\>=\sum_j\alpha_j|S_j\>$.  Apart from an irrelevant phase factor,
this state is uniquely determined by the POVM's normalization
condition $\int d\phi\:\Pi_\phi=\openone$ (see \cite{chiara}). Indeed,
this condition is satisfied only if $|\alpha_j|=1/\sqrt{N+1}$ for all
$j$.  Hence an optimal POVM is
\begin{eqnarray}
\Pi_\phi=(N+1)|\phi\>\<\phi|,\mbox{ with }
|\phi\>\equiv\tfrac1{\sqrt{N+1}}\sum_{n=0}^Ne^{i2\pi n\phi}|S_n\>.
\labell{povmo}\;
\end{eqnarray}
Then, the conditional probability of finding the result $\phi$ (which
is our estimate of the unknown parameter) when the true value is
$\varphi$ is
\begin{eqnarray}
&&p(\phi|\varphi)=(\<\varphi|^{\otimes N})\Pi_\phi(|\varphi\>^{\otimes
  N})\\&&=
\sum_{n,n'=0}^N\frac1{2^N}\sqrt{\left(\begin{matrix}N\cr
      n\end{matrix}\right)
\left(\begin{matrix}N\cr
    n'\end{matrix}\right)}\:{e^{i2\pi (\phi-\varphi)(n-n')}}
\labell{condprob}\;,
\end{eqnarray}
whence one can calculate the mutual information $I(\phi:\varphi)$. Its
asymptotic scaling (supplementary material) is \begin{align}
  I(\phi:\varphi)\to\tfrac12\log_2N+\tfrac12\log_2{\tfrac{2\pi}e}
\simeq  \tfrac 12\log_2N+0.6
\labell{mutui}\;,
\end{align}
namely the standard
quantum limit for the mutual information (apart from a small additive
constant). The explicit evaluation of $I(m:\varphi)$ shows that it
quickly attains the asymptotic expression, Fig.~\ref{f:numeric}a.
This proves that separable probes can achieve at most the standard
quantum limit.

The above strategy uses separable input states, but an entangled POVM
$\Pi_\phi$ (the states $|\phi\>$ are entangled). We now show that the
standard quantum limit can be achieved also by a strategy separable
both at the input and at the measurement. Indeed, consider the
strategy in which we measure the separable state $|\varphi\>^{\otimes
  N}$ with a projective POVM which projects onto the states
$|\pm\>\propto|0\>\pm|1\>$ each of the $N$ qubits separately. The
outcome will be a string $\vec m$ of $N$ zero-one results
corresponding to outcome ``$+$'' or ``$-$'' at each qubit
respectively.  The probability of each outcome is
$p(+|\varphi)=\cos^2(\pi\varphi)$ and
$p(-|\varphi)=\sin^2(\pi\varphi)$, so the probability of obtaining
the whole string $\vec m$ is
\begin{eqnarray}
&&p(\vec m|\varphi)=
\sin^{2\kappa}(\pi\varphi)\cos^{2(N-\kappa)}(\pi\varphi)
\labell{probs}\;,
\end{eqnarray}
where $\kappa$ is the number of ones in the string $\vec m$ (its
Hamming weight).  The unknown parameter is easily estimated from the
vector $\vec m$ as $\kappa/N$. The marginal probability of the
string $\vec m$ is then
\begin{align}
p(\vec m)=\int_0^{1}{d\varphi}\:p(\vec m|\varphi)=
\frac{(2(N-\kappa))!(2\kappa)!}{2^{2N}(N-\kappa)!\kappa!N!}
\labell{marg}\;.
\end{align}
Whence mutual information is (supplementary material)
\begin{eqnarray}
&&I(\vec
m:\varphi)=N/\ln
  2\labell{mt}
\\&&\nonumber
+\sum_{\kappa=0}^N\frac{(2(N-\kappa))!(2\kappa)!}
{4^N((N-\kappa)!)^2(\kappa!)^2}\log_2\frac{(N-\kappa)!\kappa!N!}
{(2(N-\kappa))!(2\kappa)!}
\;.
\end{eqnarray}
The asymptotic scaling of $I(\vec m:\varphi)$ of Eq.~\eqref{mt} for
large $N$ was numerically checked (Fig.~\ref{f:numeric}b) and goes as
$\simeq\log(N)/2-0.395$ (the constant was evaluated numerically), as
expected from the standard quantum limit.

\sectiona{Beyond qubits} We now drop the assumption of two-dimensional
probes (qubits) and consider the effect of a $d$-dimensional Hilbert
space of the probes.  In this case, we must consider the
transformation $U_\varphi=\sum_{n=0}^{d-1}e^{i2\pi n\varphi}|n\>\<n|$,
where $|n\>$ are eigenstates with eigenvalue $n$ of the generator $H$
of $U_\varphi$.  Intuitively, one expects that a two-dimensional probe
will give outcomes in bits (base-2 numbers) and that a $d$-dimensional
probe will give outcomes in base-$d$ numbers. We will see that this
intuition is correct: one will asymptotically gain the factor
$\log_2d$ of a change of basis in the logarithms in the mutual
information definition. We can prove this result using a
$d$-dimensional extension of the QPEA for the sequential and entangled
protocols, and using the Pegg-Barnett states for the separable
protocol (also shown in \cite{nt}).

The QPEA for $d$-dimensional systems \cite{qpead} is a straightforward
extension of the QPEA. Its output is a number $m$ composed of $t$
base-$d$ digits, whence the parameter $\varphi$ can be estimated as
$m/d^t$. The conditional probability of obtaining $m$ given $\varphi$
is
\begin{align}
  p(m|\varphi)=\frac{\sin^2(\pi\varphi d^{t})}{d^{2t}
    \sin^2[\pi(\varphi-m/d^{t})]}
\labell{profd}\;,
\end{align} analogous to \eqref{probqpea}.  The mutual
information is then
\begin{eqnarray}
  &&
  I(m:\varphi)=t\log_2d+\int_0^1
{d\varphi}\sum_mp(m|\varphi)\log_2p(m|\varphi)
  \nonumber\\&&
  \to 
  t\log_2d-1.2199
\labell{mutinfod}\;,
\end{eqnarray}
where the asymptotic scaling is derived in the same way as for
\eqref{mmm}. The ${t}\sim\log_2N$ factor in \eqref{mutinfod} accounts
for the Heisenberg scaling of the QPEA, while the $\log_2d$ term
accounts for the increase in the dimensionality of the probes. The
form of $I(m:\varphi)$ as a function of $d$ is the same as the one
shown in Fig.~\ref{f:mutinf} if one replaces $N+1$ with $d^t$: compare
\eqref{profd} with \eqref{probqpea}. Hence as in the previous case,
the asymptotic scaling \eqref{mutinfod} kicks in very rapidly.

\togli{
\begin{figure}[hbt]
\begin{center}
\epsfxsize=.95\hsize\leavevmode\epsffile{dim.eps}
\end{center}
\vspace{-.5cm}
\caption{\comment{This figure can be removed, it's basically the same
    graph as Fig.~\ref{f:mutinf}}Effect of the probe dimension $d$.
  (a)~Plot of the mutual information $I(j:\phi)$ of
  Eq.~\eqref{mutinfod} as a function of $d$ (blue) and of the function
  $\log_2(d)$ (light blue). (b)~Plot of the ratio
  $I(j:\phi)/(\log_2(d)-1.2199)$ as a function of $d$: notice the
  convergence to $1$ of the ratio.  }
\label{f:qudits}\end{figure}}

In the separable case, we can find a similar $\log_2d$ factor by
preparing each $d$-dimensional probe in the Pegg-Barnett state
$\sum_n|n\>/\sqrt{d}$ \cite{pb}, which is evolved by $U_\varphi$ into
the state $|\varphi,d\>\equiv\sum_ne^{2\pi in\varphi}|n\>/\sqrt{d}$. A
measurement that extracts information from the probe asymptotically
approaching $\log_2d$ bits is a projective POVM onto the states
$|\varphi_j,d\>$ with $\varphi_j=j/d$ (with $j=0,\cdots,d-1$)
\cite{nt}.  This is equivalent to the above $d$-dimensional QPEA for a
single probe ${t}=1$, so the mutual information of this Pegg-Barnett
procedure is given by Eq.~\eqref{mutinfod} with ${t}=1$, where again
we find a $\log_2d$ factor. Hence, also in the separable case, an
increase in the probes dimension leads to a $\log_2d$ increase in the
estimation precision.

Note that, also in the RMSE case, an increase in the probe dimension
increases the precision, because we can access larger eigenvalues of
the generator of $U_\varphi$. However, in that case, one can always
restrict the probes to a two-dimensional subspace, spanned by the
eigenvectors $|0\>$ and $|d-1\>$ relative to the minimum and maximum
eigenvalues of the generator $H$ \cite{qmetr}. In the mutual-info
case, this is not true anymore: the above $\log_2d$ increase in
precision is absent if we limit the probe states to the subspace
spanned by these two states (supplementary material). Interestingly,
the Pegg-Barnett states are known to be useless in achieving the RMSE
base Heisenberg scaling in the dimension $d$ \cite{nonp}, in contrast
to the above $\log_2d$ scaling result.  These two facts emphasize
that, although RMSE and mutual-info give consistent indications on the
measurement precision, they really capture different aspects of it.

\sectiona{Conclusions}
In conclusion, we have given an information-theoretic version of
quantum metrology, leading to the main results of ordinary RMSE-based
quantum metrology, but highlighting some peculiar differences from it.
We did not consider the effect of noise and experimental imperfections
here, leaving it to future work, since this substantially complicates
the theory, as happens in the RMSE-case,
e.g.~\cite{chiara1,rafalguta,davidov,rafal,kavan,ali}.

\subsection*{Acknowledgments}
LM acknowledges the FQXi foundation for funding. MH thanks
A.T.~Rezakhani for support and P.~Perinotti for useful discussions,
the MSRT of Iran and Iran Science Elites Federation for funding, and
the University of Pavia for hospitality.

\vfill \pagebreak \onecolumngrid

\setcounter{equation}{0}
\renewcommand{\theequation}{S\arabic{equation}}
\section*{\Large Supplementary Material}
\section{The Heisenberg bound is $\simeq\log_{2}N$} 
We prove that the Heisenberg bound is asymptotically equal to $\log_2N$ for large $N$. This follows from the proof that $N$ applications of the unitary $U_\varphi$ for unknown $\varphi$ cannot increase the entropy beyond $\log_2(N+1)$, whenever the initial probe state $\rho_0$ is pure. Consider the projector $P_k$ that projects onto an $N$-qubit state with $k$ ones, e.g.~for $N=3$ $P_0=\ketbra{000}{000}$, $P_1=\ketbra{001}{001}+\ketbra{010}{010}+\ketbra{100}{100}$, $P_2=\ketbra{110}{110}+\ketbra{101}{101}+\ketbra{011}{011}$, $P_3=\ketbra{111}{111}$. The set $\{P_k\}$ is a projective POVM, namely $P_k^2=P_k$ and $\sum_{k=0}^NP_k=\openone$.  Use this POVM to build the following chain of inequalities 
\begin{align}\label{entropy}
& S(\sum_\varphi p_\varphi\rho_\varphi)= S(\sum_\varphi p_\varphi 
  U^{\otimes N}_\varphi\rho_0{U^{\otimes N}_\varphi}^\dag) 
\leqslant S(\sum_{k=0}^NP_k\sum_\varphi p_\varphi 
  U^{\otimes N}_\varphi\rho_0{U^{\otimes N}_\varphi}^\dag
  P_k)
=S(\sum_{k=0}^N P_k \rho_0 P_k)\leqslant\log_{2}{(N+1)},
\end{align}
where the first inequality follows from the fact that a projective
measurement increases entropy, the following equality follows from the
fact that the projection removes any $\varphi$ dependence, and the
last inequality follows from the fact that there are $N+1$ terms in
the sum over $k$. A similar proof holds also if we apply sequentially
the $N$ unitaries $U_\varphi$ to a single probe or for hybrid
sequential/parallel strategies.\\
We are interested in the asymptotic scaling for large $N$, so
$\log_2(N+1)\simeq\log_2N$, and we consider this last as the
Heisenberg bound scaling, since joining Eq.~\eqref{entropy} with the
Holevo bound, we find that
\begin{align}\label{hol} 
I(\vec{m}:\varphi)\leqslant\log_2(N+1)\simeq\log_{2}{N}. 
\end{align} 
\section{the mutual information of the QPEA is $\simeq\log_{2}(N+1)-1.2199$ }
\begin{figure*} 
\includegraphics[width=0.4\linewidth]{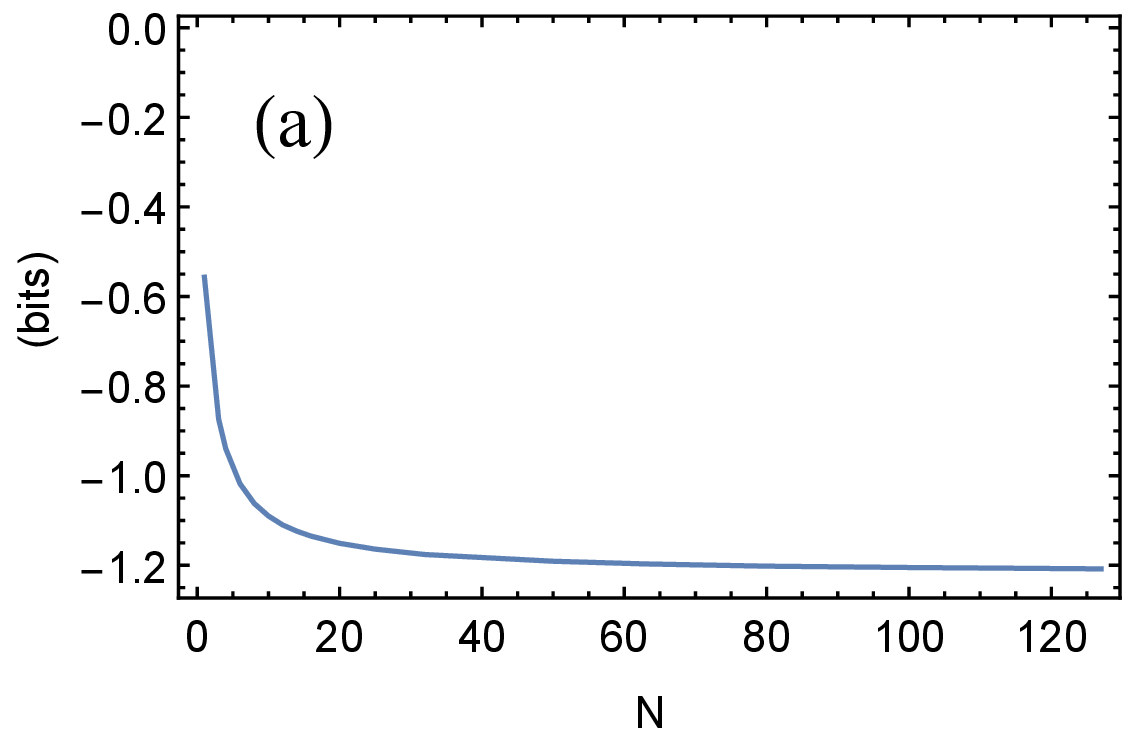}\quad\includegraphics[width=0.4\linewidth]{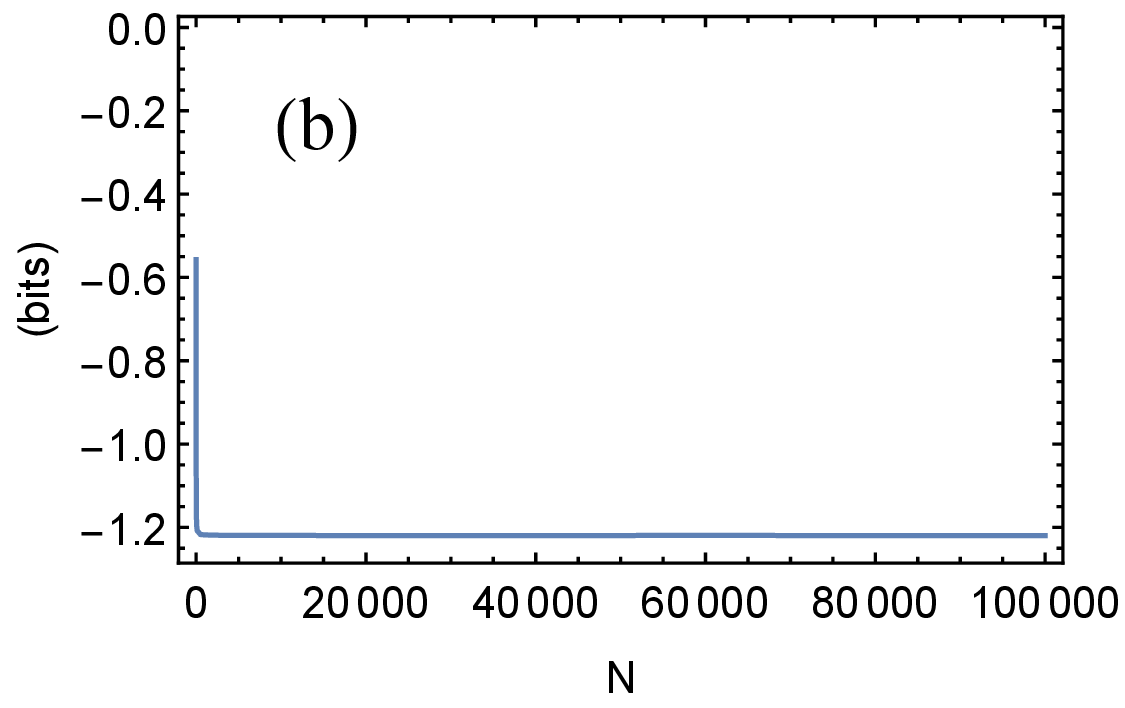}
\caption{Explicit evaluation of the quantity in Eq.~(\ref{muQQn}) as a function of $ N $. (a) Plot up to $ N=127 (t=7) $. (b) Plot up to $ N=100000 $. It shows that Eq.~(\ref{muQQn}) is equal to $ -1.299 $ for large $ N $.} 
\label{fig1} 
\end{figure*}

We prove that the mutual information of the QPEA is asymptotically
equal to $\log_{2}N-1.2199$ for large $N$. By using Eq.(2) of the main
text, we have
\begin{align}\label{muQQ}
I(m:\varphi)&=\sum_{m=0}^{N}{\int_{0}^{1}{d\varphi ~p(m|\varphi)p(\varphi)\log_{2}\left[\frac{p(m|\varphi)}{p(m)} \right]}}\nonumber\\
&=\log_{2}(N+1)+\sum_{m=0}^{N}{\int_{0}^{1}{d\varphi ~\frac{1}{(N+1)^{2}}\frac{\sin ^{2}((N+1)\pi\varphi)}{\sin ^{2}(\pi(\varphi-\frac{m}{(N+1)}))}\log_{2}\left[\frac{1}{(N+1)^{2}} \frac{\sin ^{2}((N+1)\pi\varphi)}{\sin ^{2}(\pi(\varphi-\frac{m}{(N+1)}))}\right] 
}}\nonumber\\
&=\log_{2}(N+1)+\int_{0}^{1}{d\varphi ~\log_{2}\left[\sin ^{2}((N+1)\pi\varphi)\right]}\nonumber\\
&~~~-\sum_{m=0}^{N}{\int_{0}^{1}{d\varphi ~\frac{1}{(N+1)^{2}}\frac{\sin ^{2}((N+1)\pi\varphi)}{\sin ^{2}(\pi(\varphi-\frac{m}{(N+1)}))}\log_{2}\left[(N+1)^{2}\sin ^{2}(\pi(\varphi-\frac{m}{N+1}))\right] 
}}.
\end{align}
The second term of Eq.~(\ref{muQQ}) is equal to $ -2 $ for all $ N $. We focus on the third term of Eq.~(\ref{muQQ}). First we can get rid of the sum over $ m $ by noticing that $ \sin ^{2}((N+1)\pi\varphi)=\sin ^{2}((N+1)\pi(\varphi-\frac{m}{N+1})) $ for integer $ m $. This implies that, changing the integration variable $ \varphi\rightarrow\varphi-\frac{m}{N+1} $ and using the periodicity of the $ \sin $ to restore the integration extremes, that integral can be also written as 
\begin{align}
-\sum_{m=0}^{N}{\frac{1}{(N+1)^{2}}\int_{0}^{1}{d\varphi ~\frac{\sin
      ^{2}((N+1)\pi\varphi)}{\sin ^{2}(\pi\varphi)}\log_{2}\left[
      (N+1)^{2}\sin ^{2}(\pi\varphi)\right]}}\\
=-\frac{1}{(N+1)}\int_{0}^{1}{d\varphi ~\frac{\sin ^{2}((N+1)\pi\varphi)}{\sin ^{2}(\pi\varphi)}\log_{2}\left[ (N+1)^{2}\sin ^{2}(\pi\varphi)\right]}.
\label{muQQ3}\end{align}
Now we can change the integration variable to $ x=(N+1)\pi\varphi $ to write Eq.~(\ref{muQQ3}) as 
\begin{equation}
-\frac{1}{(N+1)}\int_{0}^{\pi (N+1)}{\frac{dx}{\pi (N+1)}\frac{\sin ^{2}x}{\sin ^{2}(\frac{x}{(N+1)})}\log_{2}\left[ (N+1)^{2}\sin ^{2}(\frac{x}{N+1})\right]}.
\end{equation}
It is clear that the major contributions to this integral come from
the regions where the $ \sin ^{2}(\frac{x}{(N+1)}) $ in the integrand
is null, namely for $ x\rightarrow 0^{+} $ and $ x\rightarrow\pi
(N+1)^{-} $. Indeed the integrand has a logarithmic divergence there.
To look at the asymptotic behavior for $ (N+1)\rightarrow\infty $, it
is better to move this region entirely in the vicinity of $
x\rightarrow 0 $ by using the periodicity of the integrand to change
the integration extremes from $ \int_{0}^{\pi (N+1)}dx$ to $\int_{-\frac{\pi
    (N+1)}{2}}^{\frac{\pi (N+1)}{2}}dy $. Expanding to first order in
$ \frac{y}{(N+1)} $ the integrand (which is a good expansion in the
region of interest $y\rightarrow 0$), we get
\begin{align}
&-\int_{-\frac{\pi (N+1)}{2}}^{\frac{\pi (N+1)}{2}}{\frac{dy}{\pi}\frac{\sin^{2}y}{y^{2}}\log_{2}\left[ y^{2}\right]}
\rightarrow -\int_{-\infty}^{\infty}{\frac{dy}{\pi}\frac{\sin^{2}y}{y^{2}}\log_{2}\left[ y^{2}\right] }=2\left(\frac{\gamma+\ln{2}-1}{\ln{2}}\right)\simeq 0.7801,
\end{align}
where $ \gamma $ is the Euler-Mascheroni constant. Hence, we proved
\begin{align}\label{muQQn}
\sum_{m=0}^{N}{\int_{0}^{1}{d\varphi\frac{1}{(N+1)^{2}}\frac{\sin ^{2}((N+1)\pi\varphi)}{\sin ^{2}(\pi(\varphi-\frac{m}{(N+1)}))}\log_{2}\left[\frac{1}{(N+1)^{2}} \frac{\sin ^{2}((N+1)\pi\varphi)}{\sin ^{2}(\pi(\varphi-\frac{m}{(N+1)}))}\right] 
}}\rightarrow -2+2\left(\frac{\gamma+\ln{2}-1}{\ln{2}}\right)\simeq -1.2199.
\end{align}
Also the numerical calculations shows that Eq.~(\ref{muQQn}) converges to $ -1.299 $ for large $ N $, see Fig.~(\ref{fig1}). For the QPEA $ I(m,\varphi)\rightarrow\log_{2}{N}-1.299. $  
\section{the mutual information of the QPEA with a separable detection}
We show that, if one uses the single-qubit measurement which projects
each probe onto $ \ket{\pm} $ for the QPEA, the Heisenberg bound is
\textit{not} achieved. Consider the parallel QPEA (Fig.~1c). The state
of j-th probe after the transformations $ U_{\varphi} $ is $
\frac{1}{\sqrt{2}}(\ket{0}^{2^{j}}+e^{i2\pi\varphi
  2^{j}}\ket{1}^{2^{j}}) $ (with $ j=0,1,...,t-1 $). One can measure
each qubit of the probe separately with the projective operators which
project onto $ \ket{\pm}\propto \ket{0}\pm\ket{1} $. Therefore the
outcome will be string $ \vec{m} $ of $ 2^{j} $ bits corresponding to
outcome ``{$ + $}'' or ``{$ - $}'' at each qubit respectively. If the
number of ones in the string (its Hamming weight) is even (odd), the
probability $ p(\vec{m}|\varphi) $ is $
\frac{1}{2^{(2^{j}-1)}}\cos^{2}(\pi\varphi 2^{j}) \left(
  \frac{1}{2^{(2^{j}-1)}}\sin^{2}(\pi\varphi 2^{j}) \right) $. Hence
the mutual information of j-th probe is
\begin{align}
I^{(j)}(\vec{m}:\varphi)&=\log_{2}{2^{(2^{j})}}+\int_{0}^{1}{d\varphi ~\cos^{2}(\pi\varphi 2^{j})\log_{2}\left[ \frac{1}{2^{(2^{j}-1)}}\cos^{2}(\pi\varphi 2^{j})\right]+\sin^{2}(\pi\varphi 2^{j})\log_{2}\left[ \frac{1}{2^{(2^{j}-1)}}\sin^{2}(\pi\varphi 2^{j})\right] }\nonumber\\
&=1+\int_{0}^{1}{d\varphi ~\cos^{2}(\pi\varphi 2^{j})\log_{2}\left[\cos^{2}(\pi\varphi 2^{j})\right]+\sin^{2}(\pi\varphi 2^{j})\log_{2}\left[\sin^{2}(\pi\varphi 2^{j})\right]}\nonumber\\
&=1+\log_{2}{\frac{e}{4}}\simeq 0.44~.
\end{align}
The total mutual information of $ t $ probes is $ 0.44~t=0.44\log_{2}(N+1) $ and the Heisenberg bound is not achieved. We cannot even attain the standard quantum limit $ \frac{1}{2}\log_{2}N $. Since this is the same result that is obtained from the optimal POVM of the Davies theorem, we can conclude that this POVM is also an optimal one. Then, using Davies' POVM separately on each qubit would not give any advantage over the above calculation.
\section{the mutual information of the separable parallel strategy with an entangled measurement}
We show that the mutual information of the parallel strategy with a separable initial state and entangled POVM is asymptotically equal to $ \frac{1}{2}\log_{2}N+0.6 $. The conditional probability for the measurement described by the POVM $ \Pi_{\phi} $ of Eqs.~(5,6) of the main text is  
\begin{align}
p(\phi|\varphi)&=(\bra{\varphi}^{\otimes
  N})\Pi_{\phi}(\ket{\varphi}^{\otimes
  N})=\sum_{n^{\prime},n=0}^{N}{\sqrt{\frac{1}{2^N}\binom{N}{n^{\prime}}}{\sqrt{\frac{1}{2^N}\binom{N}{n}}}}\exp\left[i2\pi(\phi -\varphi)(n^{\prime}-n) \right]. \label{CQt1}
\end{align}
By using the following approximation that is valid for $ N\gg1 $:
\begin{equation}\label{binexp}
\binom{N}{n}\simeq 2^{N}\sqrt{\frac{2}{\pi N}}~\exp\left[ -\frac{(n-\frac{N}{2})^{2}}{\frac{N}{2}}\right],
\end{equation}
in Eq.~(\ref{CQt1}), we have 
\begin{align}{\label{conCQ1}}
p(\phi|\varphi)&\simeq\sqrt{\frac{2}{\pi N}}\sum_{n^{\prime},n=0}^{N}{\exp\left[ -\frac{(n^{\prime}-\frac{N}{2})^{2}}{N}\right]\exp\left[ -\frac{(n-\frac{N}{2})^{2}}{N}\right]}\exp\left[i2\pi(\phi -\varphi)(n^{\prime}-n) \right]\nonumber\\
&=\sqrt{\frac{2}{\pi N}}\left( \sum_{n^{\prime}=0}^{N}{\exp\left[ -\frac{(n^{\prime}-\frac{N}{2})^{2}}{N}\right]\exp\left[-i2\pi(\phi -\varphi)(n^{\prime}-\frac{N}{2})\right]}\right)\nonumber\\&\times\left( \sum_{n=0}^{N}{\exp\left[ -\frac{(n-\frac{N}{2})^{2}}{N}\right]\exp\left[i2\pi(\phi -\varphi)(n-\frac{N}{2}) \right]}\right).
\end{align}
Considering $ n^{\prime} $ and $ n $ as continuous variables and replacing the sum with an integral we obtain
\begin{align}{\label{CQt2}}
p(\phi|\varphi)&\simeq\sqrt{2\pi N}~\exp\left[ -2\pi ^{2}(\phi -\varphi)^{2}N\right].
\end{align}
Substitute Eq.~(\ref{CQt2}) in the mutual information relation (using uniform prior $ p(\phi)=1 $) 
\begin{align}\label{CQt3}
I(\phi :\varphi)&=\int_{0}^{1}{\int_{0}^{1}{d\varphi ~d\phi ~p(\phi |\varphi)\log_{2}\left[p(\phi |\varphi) \right]}}\nonumber\\
&\simeq\int_{0}^{1}{\int_{0}^{1}{d\varphi ~d\phi ~\sqrt{2\pi N}~\exp\left[-2\pi ^{2}(\phi -\varphi)^{2}N\right]\log_{2}\left[\sqrt{2\pi N}~\exp\left[ -2\pi ^{2}(\phi -\varphi)^{2}N\right]\right]}}\nonumber\\
&=\frac{1}{2}\log_{2}{N}+\log_{2}{\sqrt{2\pi}}-(2\pi^{2}N)\left(\log_{2}e \right)\int_{0}^{1}{\int_{0}^{1}{d\varphi ~d\phi ~\sqrt{2\pi N}~(\phi -\varphi)^{2}\exp\left[-2\pi ^{2}(\phi -\varphi)^{2}N\right]}}.
\end{align}
Consider the third term of Eq.~(\ref{CQt3}) and define $\omega:=2\pi^{2}N $,
\begin{align}\label{3CQ}
  &-\sqrt{2\pi N}\omega\left(\log_{2}e \right)\int_{0}^{1}{\int_{0}^{1}{d\varphi ~d\phi ~(\phi -\varphi)^{2}\exp\left[-\omega(\phi -\varphi)^{2}\right]}}\nonumber\\
  &=-\sqrt{2\pi N}\omega\left(\log_{2}e
  \right)\int_{0}^{1}{\int_{0}^{1}{d\varphi ~d\phi
      \left[-\frac{d}{d\omega}\left( \exp\left[-\omega(\phi
            -\varphi)^{2}\right]\right)\right]}}\nonumber\\
&=\sqrt{2\pi N}\omega\left(\log_{2}e
\right)\frac{d}{d\omega}\left[\frac{-1+e^{-\omega}+\sqrt{\pi\omega}\erf(\sqrt{\omega})}{\omega}
\right] \nonumber\\ 
  &=\sqrt{2\pi N}\left(\log_{2}e
  \right)\left[-\frac{1}{2}\sqrt{\frac{\pi}{\omega}}\erf(\sqrt{\omega})+\frac{1}{\omega}(1-e^{-\omega})\right] \nonumber\\ 
  &\simeq -\frac{1}{2}\log_{2}{e}~~~~~~~~~~~~~ N\gg 1,
\end{align}
where $ \erf(x):=\frac{1}{\sqrt{\pi}}\int_{-x}^{x}dt\:{e^{-t^{2}}} $
denotes the error function. This function is equal to $ 1 $ for large
$ x $. Replacing Eq.~(\ref{3CQ}) in Eq.~(\ref{CQt3}) 
\begin{align}\label{conCQ3}
I(\phi
:\varphi)&\rightarrow\frac{1}{2}\log_{2}{N}+\log_{2}{\sqrt{\frac{2\pi}{e}}}
\simeq\frac{1}{2}\log_{2}{N}+0.6~~~~~~~~~~~~~~ N\gg 1.
\end{align}
\section{the mutual information of the parallel separable strategy with a separable measurement}
\begin{figure*} 
\includegraphics[width=0.4\linewidth]{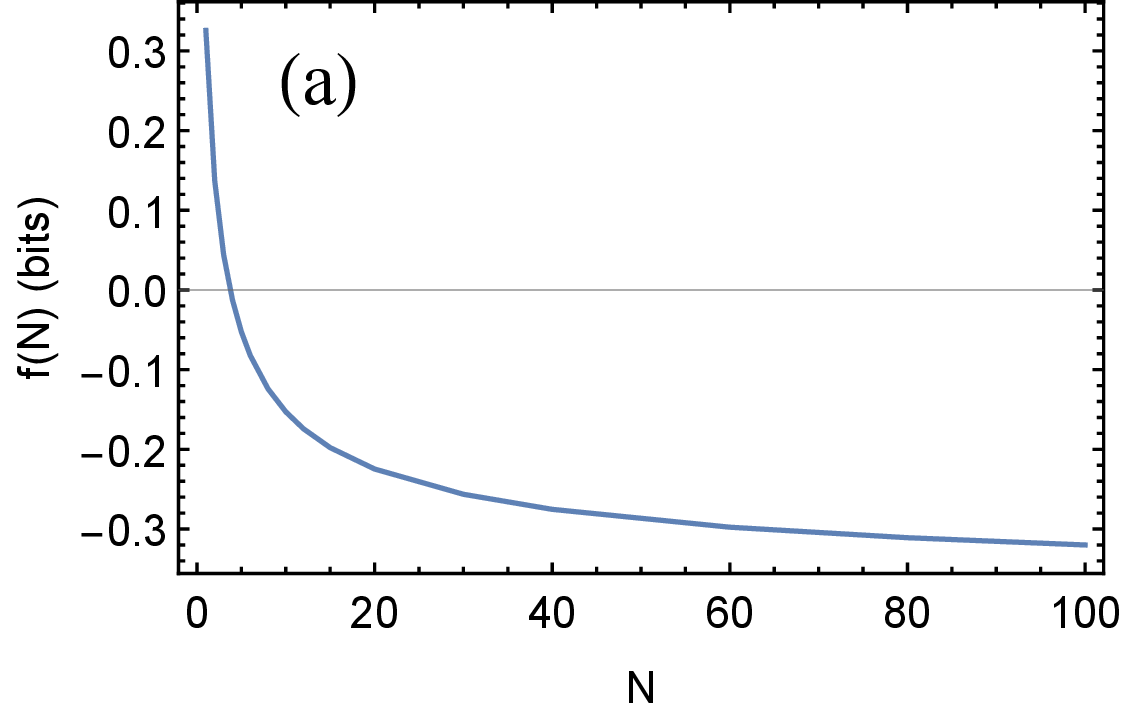}\quad\includegraphics[width=0.4\linewidth]{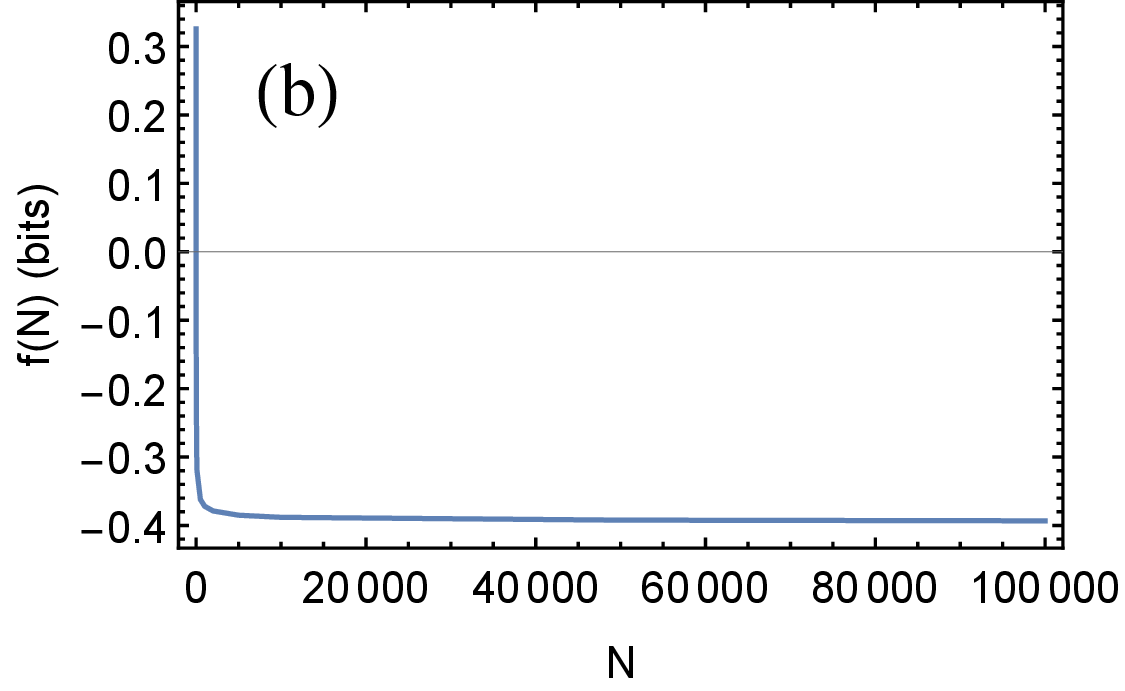}
\caption{Numerical evaluation of $ f(N) $ up to (a) $ N=100 $ and (b) $ N=100000 $ .} 
\label{fig2} 
\end{figure*}

We show that the mutual information of the parallel strategy with a separable initial state and separable POVM is asymptotically equal to $ \frac{1}{2}\log_{2}N-0.395 $. We want to derive Eq.~(12) of the main text. By using Eqs.~(10,11) of the main text, we have:
\begin{align}\label{CC2p}
I(\vec{m}:\varphi)&=\sum_{\kappa=0}^{N}{{\int_{0}^{1}{d\varphi\binom{N}{\kappa}\left(\cos^{2}(\pi\varphi) \right) ^{N-\kappa}\left(\sin^{2}(\pi\varphi) \right) ^{\kappa}\log_{2}\left[ \left(\cos^{2}(\pi\varphi) \right) ^{N-\kappa}\left(\sin^{2}(\pi\varphi) \right) ^{\kappa}\right]}}}\nonumber\\
&~~~+\sum_{\kappa=0}^{N}{\int_{0}^{1}{d\varphi\binom{N}{\kappa}\left(\cos^{2}(\pi\varphi) \right) ^{N-\kappa}\left(\sin^{2}(\pi\varphi) \right) ^{\kappa}\log_{2}\left[ {\frac{4^{N}(N-\kappa)!\kappa !N!}{(2(N-\kappa))!(2\kappa)!}}\right] }}.
\end{align} 
The first part of Eq.~(\ref{CC2p}) is equal to
\begin{align}\label{CCp1}
&\sum_{\kappa=0}^{N}{{\int_{0}^{1}{\binom{N}{\kappa}\left(\cos^{2}(\pi\varphi) \right) ^{N-\kappa}\left(\sin^{2}(\pi\varphi) \right) ^{\kappa}({N-\kappa})\log_{2}\left[\cos^{2}(\pi\varphi)\right] }d\varphi}}\nonumber\\
&+\sum_{\kappa=0}^{N}{{\int_{0}^{1}{\binom{N}{\kappa}\left(\cos^{2}(\pi\varphi) \right) ^{N-\kappa}\left(\sin^{2}(\pi\varphi) \right) ^{\kappa}\kappa\log_{2}\left[\sin^{2}(\pi\varphi)\right] }d\varphi}},\nonumber\\
&=\int_{0}^{1}\log_{2}\left[\cos^{2}(\pi\varphi)\right]\sum_{\kappa=0}^{N}{\binom{N}{\kappa}\left(\cos^{2}(\pi\varphi) \right) ^{N-\kappa}\left(\sin^{2}(\pi\varphi) \right) ^{\kappa}({N-\kappa})}~d\varphi\nonumber\\
&~~+\int_{0}^{1}\log_{2}\left[\sin^{2}(\pi\varphi)\right]\sum_{\kappa=0}^{N}{\binom{N}{\kappa}\left(\cos^{2}(\pi\varphi) \right) ^{N-\kappa}\left(\sin^{2}(\pi\varphi) \right)^{\kappa} \kappa}~d\varphi\nonumber\\
&=N\int_{0}^{1}{\left\lbrace\left(\cos^{2}(\pi\varphi) \right)\log_{2}\left[\cos^{2}(\pi\varphi)\right]+\sin^{2}(\pi\varphi)\log_{2}\left[ \sin^{2}(\pi\varphi)\right]\right\rbrace d\varphi},\nonumber\\
&=N\left(\log_{2}{\frac{e}{4}}\right),
\end{align}
where the second equality uses the moment-generating function method.\\
To calculate the integral in the second line of Eq.~\eqref{CC2p}
consider the definition of gamma function as $
\Gamma(z):=\int_{0}^{\infty}{t^{z-1}e^{-t}dt} $, so we have
\begin{align}\label{gam}
\Gamma(z) \Gamma(s):=\int_{0}^{\infty}{\int_{0}^{\infty}{t^{z-1}u^{s-1}e^{-(t+s)}~dt~du}}.
\end{align}
We define $ t:=x^{2} $, $ u:=y^{2} $ and replace them in Eq.~(\ref{gam})
\begin{align}
\Gamma(z) \Gamma(s):=4\int_{0}^{\infty}{\int_{0}^{\infty}{x^{2z-1}y^{2s-1}e^{-(x^{2}+y^{2})}~dx~dy}}.
\end{align}
For simplicity, we use the polar coordinate so $ x=r\cos\theta,y=r\sin\theta $
\begin{align}\label{gam1}
\Gamma(z)\Gamma(s):&=4\int_{0}^{\frac{\pi}{2}}{d\theta ~\cos^{2z-1}\theta\sin^{2s-1}\theta} 
\int_{0}^{\infty}{dr~r^{2(z+s)-2}~re^{-r^{2}}}%
\nonumber\\&
=4\left( \int_{0}^{\frac{\pi}{2}}{d\theta ~\cos^{2z-1}\theta\sin^{2s-1}\theta}\right)  
\frac{1}{2}\left( \int_{0}^{\infty}{2rdr~(\underbrace{r^{2}}_{r^{\prime}})^{(z+s)-1}e^{-r^{2}}}\right)\nonumber\\
&=2\left( \int_{0}^{\frac{\pi}{2}}{d\theta ~\cos^{2z-1}\theta\sin^{2s-1}\theta }\right)  
\left(
  \int_{0}^{\infty}{d{r^{\prime}}{r^{\prime}}^{(z+s)-1}e^{r^{\prime}}}\right)
=2\left( \int_{0}^{\frac{\pi}{2}}{d\theta ~\cos^{2z-1}\theta\sin^{2s-1}\theta}\right)  
\Gamma(z+s),
\end{align}
Therefore
\begin{equation}\label{intbeta}
\int_{0}^{\frac{\pi}{2}}{d\theta ~(\cos^{2}\theta)^{z-\frac{1}{2}}(\sin^{2}\theta)^{s-\frac{1}{2}}}=\frac{1}{2}\frac{\Gamma(z)\Gamma(s)}{\Gamma(z+s)}.
\end{equation}
 For non-negative integer value of $ a $, we have $
 \Gamma(a+\frac{1}{2})=\frac{(2a)!}{4^{a}a!}\sqrt{\pi} $. By
 substituting Eq.~(\ref{intbeta}) in the second line  of Eq.~(\ref{CC2p}), we get
\begin{align}\label{CCp21}
\sum_{\kappa=0}^{N}{\frac{(2(N-\kappa))!(2\kappa)!}{4^{N}((N-\kappa)!)^{2}(\kappa !)^{2}}\log_{2}\left[ {\frac{4^{N}(N-\kappa)!\kappa !N!}{(2(N-\kappa))!(2\kappa)!}}\right] }\simeq & 2N+\frac{1}{2}\log_{2}N+\log_{2}{\sqrt{2\pi}}+N\log_{2}N-N\log_{2}e\nonumber\\
&+\sum_{\kappa=0}^{N}{\frac{(2(N-\kappa))!(2\kappa)!}{4^{N}((N-\kappa)!)^{2}(\kappa !)^{2}}\log_{2}\left[ {\frac{(N-\kappa)!\kappa !}{(2(N-\kappa))!(2\kappa)!}}\right] },
\end{align}
where we use the Stirling's formula ($ \nu
!\sim\sqrt{2\pi\nu}(\frac{\nu}{e})^{\nu} $ for large $ \nu $). By
substituting Eq.~(\ref{CCp1}) and Eq.~(\ref{CCp21}) in Eq.~(\ref{CC2p}), we get
\begin{align}\label{CC2pe}
I&\simeq\frac{1}{2}\log_{2}N+\underbrace{\log_{2}{\sqrt{2\pi}}+N\log_{2}N+\sum_{\kappa=0}^{N}{\frac{(2(N-\kappa))!(2\kappa)!}{4^{N}((N-\kappa)!)^{2}(\kappa !)^{2}}\log_{2}\left[ {\frac{(N-\kappa)!\kappa !}{(2(N-\kappa))!(2\kappa)!}}\right] }}_{f(N)},\nonumber\\
&=\frac{1}{2}\log_{2}N+f(N).
\end{align}
The numerical calculation of $ f(N) $ shows that this term is asymptotically a constant and $ f(N) \rightarrow -0.395$, Fig.~(\ref{fig2}).\\
\\
\section{the mutual information for the probe in a 2-dimensional subspace}
We show that the mutual information of the probe which is spanned by the states $ \ket{0} $ and $ \ket{d-1} $ does not increase with the probe Hilbert space dimension $ d $. Suppose the initial  state is $ \ket{\psi_{0}}=\frac{1}{\sqrt{2}}(\ket{0}+\ket{d-1}) $ and the quantum state after the evolution is $ \ket{\psi_{\varphi}}=\frac{1}{\sqrt{2}}(\ket{0}+e^{i2\pi\varphi (d-1)}\ket{d-1}) $. One can use the Davies theorem to find the optimal POVM that maximizes the mutual information as $ \Pi_{\phi}=2 U_{\varphi}\ketbra{r}{r}U_{\varphi}^{\dagger} $, when restricting the Hilbert space to the subspace spanned by $ \ket{0} $ and $ \ket{d-1} $ the Davies theorem gives that $\ket{r}=\frac{1}{\sqrt{2}}(\ket{0}+\ket{d-1}) $, so the mutual information is  
\begin{align}
I(\phi :\varphi)&=\log_{2}2+2\int_{0}^{1}{\int_{0}^{1}{\cos ^{2}(\pi(\phi -\varphi)(d-1))\log_{2}\left[ \cos ^{2}(\pi(\phi -\varphi)(d-1))\right]~d\varphi}~d\phi},\nonumber\\
&=1+\log_{2}{\frac{e}{4}}\simeq 0.44~.
\end{align}
Also one can use the POVM $ \ket{\pm}\propto\ket{0}\pm\ket{d-1} $ to obtain the same result. In this case the mutual information does not depend on $ d $.


\begin{references}
\bibitem{nonp}This was proven in \cite{nt}, where they use a $\sin^2$
  window function which (for small phases) is equivalent to the
  variance.
\end{references}
\end{document}